\documentclass[twocolumn,showpacs,superscriptaddress,prb,aps,floatfix, amsmath,amssymb]{revtex4-1}

\usepackage{epsfig}
\usepackage{graphicx}
\usepackage{color}
\usepackage{soul}
\usepackage{setspace}

\begin{document}

\title{Coexistence of trapped and free excess electrons in SrTiO$_3$}

\author{Xianfeng Hao}
\affiliation{Institute of Applied Physics, Vienna University of Technology, Vienna, Austria}
\affiliation{University of Vienna, Faculty of Physics and Center for Computational Material Science, Vienna, Austria }
\affiliation{Key Laboratory of Applied Chemistry, Yanshan University, Qinhuangdao 066004, P. R. China} 

\author{Zhiming Wang}
\affiliation{Institute of Applied Physics, Vienna University of Technology, Vienna, Austria}

\author{Michael Schmid}
\affiliation{Institute of Applied Physics, Vienna University of Technology, Vienna, Austria}

\author{Ulrike Diebold}
\affiliation{Institute of Applied Physics, Vienna University of Technology, Vienna, Austria}

\author{Cesare Franchini}
\email{cesare.franchini@univie.ac.at}
\affiliation{University of Vienna, Faculty of Physics and Center for Computational Material Science, Vienna, Austria }

\date{\today}

\begin{abstract}

The question whether excess electrons in SrTiO$_3$
form free or trapped carriers is a crucial aspect for the electronic properties of this important material. This fundamental ambiguity prevents a consistent interpretation of the 
puzzling  experimental situation, where  results support one or the other scenario depending on the type of experiment that is conducted.
Using density functional theory with an on-site Coulomb interaction U, 
we show that excess electrons form small polarons if the density of electronic carriers is higher than 
$\approx$ $10^{20}\,\,\mathrm{cm}^{-3}$.
Below this value, the electrons stay delocalized or become large polarons. 
For oxygen-deficient SrTiO$_3$, small polarons confined to Ti$^{3+}$ sites are immobile at low 
temperature but can be thermally activated into a conductive state, which explains the
metal-insulator transition observed experimentally.
\end{abstract}


\maketitle 

%%%%%%%%%%%%%%%%%%%%%%%%%%%%%%%%%%%%%%%%%%%%%%%%%%%%%%
\section{INTRODUCTION}
%%%%%%%%%%%%%%%%%%%%%%%%%%%%%%%%%%%%%%%%%%%%%%%%%%%%%%

Strontium titanate is a key material in the emerging field of oxide electronics.
It exhibits many attractive properties such as superconductivity \cite{Schooley}, 
ferromagnetism \cite{Rice, Moetakef}, high thermoelectric coefficients \cite{Ohta}, blue and green light 
emission \cite{Kan}, and can accomodate a two dimensional electron gas \cite{Meevasana,Wang14}.
All these phenomena arise when a small number of electrons are introduced into the lattice.
The characteristics and mobility of excess electrons is strongly determined by their interaction
with the lattice through  electron-phonon coupling. One of the manifestations
of the electron-phonon interaction is the formation of a polaron, a quasiparticle
formed by the excess electron and the structural deformations it induces \cite{Frohlich,Devbook}.
Depending on the spatial extent of the polaronic wavefunction and associated structural distortions
two types of polarons can be identified: in the small-polaron model the range of the distortion is of 
the order of the lattice constant, whereas in the large polaron the wavefunction spreads over 
several lattice sites and exhibits a more free-carrier-like behavior.
Large polarons show a mobility inversely proportional to the number 
of longitudinal optical phonons. In contrast, small polarons form highly localized in-gap states and can hop
to adjacent lattice sites when thermally activated \cite{Dev2000}.

Experimental information on the role of polarons in SrTiO$_3$ is abundant, yet controversial 
\cite{Mazin, Devreese, Chang2010, Kohmoto, Yamada2010, Liu, Lee2011, Yamada, Ishida, Fujimori2, Higuchi, Angelo,Fujimori,LeeM2011}. 
The key question is whether it is a large or a small polaron that is formed. 
Excess electrons introduced by doping or by oxygen defects have been identified as large polarons in 
transport measurements and optical spectroscopies \cite{Mazin,Devreese}. 
The validity of this picture is supported
by recent angle-resolved photoemission spectroscopy data  \cite{Chang2010}.  
Pump-probe spectroscopy in lightly photoexcited SrTiO$_3$ suggested that itinerant electron polarons 
are the origin of diffusive lattice distortions \cite{Kohmoto}.
However, the large-polaron model cannot explain the in-gap bulk states in the conductive phase revealed in 
photoemission spectroscopy \cite{Fujimori, Fujimori2, Higuchi, Angelo}. Instead, these highly localized states 
are attributable to small polarons trapped in bulk sites. This interpretation is reinforced by the recent
observation that photoexcited electrons relax into self-trapped polaron states \cite{Yamada, Kohmoto}. 
Electron trapping has been also invoked to explain the metal-insulator transition (MIT) in 
SrTiO$_{3-\delta}$ thin films \cite{Liu} and single crystals \cite{Lee2011}. 
To complicate the situation even further, there are indications that excess electrons localized in trapped
states can coexists with free carriers \cite{Devreese, Ishida, Yamada2010, LeeM2011}.
No coherent picture can explain all these experiments so far; this has sparked an animated debate 
and poses a serious challenge to theory.

Computational studies provide additional insights. Based on density functional theory (DFT)+U, 
hybrid DFT and model Hamiltonians, it was suggested that the two additional electrons donated 
by a single oxygen vacancy V$_\mathrm{O}$ induce  one 
singly-occupied in-gap state mostly localized at the Ti atoms near the V$_\mathrm{O}$,
and one state delocalized in the conduction band \cite{Hou, Mitra, Choi, El-Mellouhi, Lin, Janotti14}. 
This conclusion successfully clarifies transport and optical properties in reduced SrTiO$_3$, 
but is inconsistent with the observed V$_\mathrm{O}$-induced MIT \cite{Liu}, because the predicted
free carriers would be mobile even at low temperature. 

The present study aims to provide a unified and coherent explanation of the role of excess electrons
in SrTiO$_3$. In particular it investigates  the conditions under which localized and free-electron-like states can form, coexist and propagate.
We show that a small-polaron regime is established if the carrier density $n$ is higher than 
a critical value, both in the low-temperature antiferrodistortive (AFD) and in the   
cubic structure. Below this critical density, the excess electrons will be delocalized or form large polarons, 
leading to a metallic solution without in-gap states, in nice agreement with experiment. 
Moreover, we provide theoretical evidence that the two excess electrons 
donated by V$_\mathrm{O}$ form two self-trapped electrons, which become mobile if thermally activated, thus supporting the temperature-driven MIT observed in reduced SrTiO$_3$ \cite{Liu}.

\section{METHODS}     
For our calculations we adopted spin-polarized DFT+U 
within the gradient-corrected Perdew-Burke-Ernzerhof (PBE) functional \cite{PBE} 
using the  VASP \cite{VASP}. An on-site $U$  was applied to the $d$ states of Ti. 
The value of the Coulomb and exchange parameters $U$ and $J$ was computed 
fully {\em ab initio} within the constrained Random Phase Approximation \cite{cRPA}.
To this aim the Kohn Sham orbitals were projected on to maximally localized Wannier 
functions \cite{Marzari1997} using the Wannier90 code. \cite{Franchini2012, Mostofi2008}
For the construction of the correlated subspace we have followed the hybrid 
"d-dp Hamiltonian"-like approach suggested by Vaugier and coworkers\cite{Vaugier},
where the Wannier functions are generated for the metal $d$ and oxygen $p$ orbitals, 
and only the $d$ bands are removed for the calculation of the effective interactions.
Following this procedure we obtained $U=4.96$ eV and $J=0.51$ eV, and set $U-J=4.5$ eV, which is
in line with previous DFT+U studies of polarons in SrTiO$_3$ \cite{Hou, Choi}. 

We considered both the cubic and the low-temperature (below 110 K) tetragonal AFD phase of SrTiO$_3$
and at the optimized lattice constants (a$_{\mathrm cubic}$ = 3.948 \AA, a$_{\mathrm AFD}$ = 3.937 \AA, 
c/a$_{\mathrm AFD}$=1.0055). Adopting these lattice constants the optmized rotation of the octahedra 
around the $z$ axis in the AFD phase is about 6$^\circ$, in agreement with previous DFT-based 
calculations \cite{wahl}. To mimic substitutional (Ti/Nb) doping and V$_\mathrm{O}$s we employed an 
$N \times N \times N$ supercell (SC) with $N=3$, 4, 5, and 6 (up to 1080 atoms). For each SC all atomic 
positions were relaxed until the residual force on each atom was less than 0.02 eV/\AA. 

The polaron formation energy is defined as the energy difference between the delocalized, free electron solution 
and the localized polaron solution. Both these solutions can be modeled by DFT.
The delocalized solution can be achieved within a standard non spin-polarized calculation.
To model the small polaron solution different schemes can be adopted as recently discussed in 
Ref. \onlinecite{Shibuya}. We have followed the following strategy, proposed by Deskins {\em et al.} \cite{Deskins}:
\begin{enumerate}
\item First, we have chosen the specific Ti sites at which we aim to stabilize a polaron
and substituted that distinct Ti with a V atom and let the structure relax within a spin-polarized
set-up. In this way we break the symmetry and allow an expansion of the oxygen octahedron around the V.
\item  Second, we placed back the Ti atom in the V position and let the structure relax within a spin-polarized
set-up by initializing a local magnetic moment equal to one in this specific Ti site.
\end{enumerate}
For particularly complicated polaronic configurations (i.e. difficult to converge) we have added an intermediate step
in which we have performed a step-2 calculation using a larger value of the on-site $U$ at the Ti site at which
we wanted to form the polaron. In this way we could more easily attract the excess electron to that specific Ti site.
In a subsequent run, we recovered the usual value of $U$ and let the system relax.
This procedure was repeated for several polaronic configurations and the corresponding total energies were
compared to the total energy of the delocalized solution.
 
We verified that the results were converged with respect to the energy cutoff (400 eV) and number of
k-points (12$\times$12$\times$12  with respect to the primitive cubic cell).
Molecular dynamics (MD) simulations of small polaron hopping in SrTiO$_{3-\delta}$ with $\delta \approx
0.00267$ were performed at simulated temperatures ranging from 700 K to 1000 K for 5 ps with a time 
step of 1 fs. The MD are also performed within the DFT+U framework. We expect that the inclusion of the inter-site 
effective interaction $V$ \cite{Campo}, which have never been employed to describe the dynamics of polarons, could
improve further the agreement with experiments. For SrTiO$_3$ our cRPA procedure delivers $V$=1.7 eV.

\section{RESULTS AND DISCUSSION}

In this section we present and discuss the results obtained for the Nb-doping and oxygen vacancies calculations.

\subsection{Nb-doping}

\begin{figure}
\includegraphics[width=1.0\columnwidth,scale=1.0]{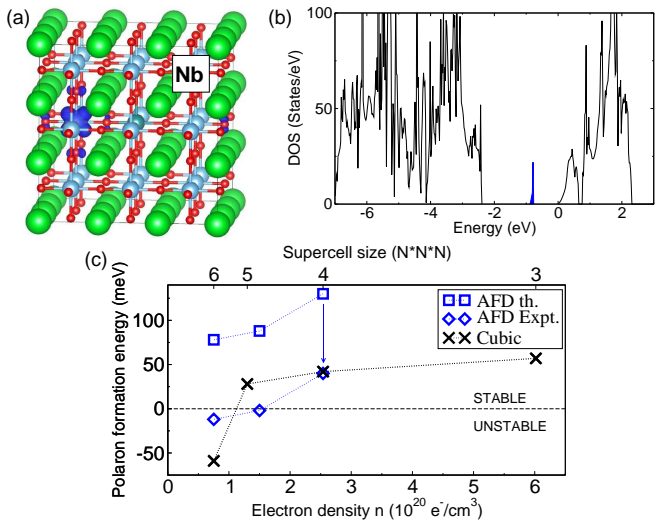}
\caption{\label{FIG:1}
(color online) Nb-doped SrTiO$_3$. (a) Isosurfaces of the spin density (dark blue;  0.045 e/{\AA}) for an $N=3$ supercell.
The green (big), light-blue/turquoise (medium size) and red (small) balls represent Sr, Ti/Nb and O atoms, 
respectively. (b) Electronic density of states (N=3); the (blue) peak in the band gap corresponds to the 
polaron state formed at a Ti$^{3+}$ site. (c) Variation of the polaron formation energy as a function of 
carrier density for the cubic 
(N=3, $6.03 \times 10^{20}$ cm$^{-3}$,  135 atoms;
 N=4, $2.54 \times 10^{20}$ cm$^{-3}$,  320 atoms; 
 N=5, $1.30 \times 10^{20}$ cm$^{-3}$,  625 atoms; 
 N=6, $0.75 \times 10^{20}$ cm$^{-3}$, 1080 atoms) 
and low-T AFD phases
($2.54 \times 10^{20}$ cm$^{-3}$, 320 atoms;
 $1.50 \times 10^{20}$ cm$^{-3}$, 540 atoms;
 $0.75 \times 10^{20}$ cm$^{-3}$, 1080 atoms) for the optimized (Th.) and experimental (Expt.) volume.
The AFD Expt. curves is obtained by shifting down the AFD Th. curve by the calculated 
energy shift at N=4, 90 meV. For the 540 atoms AFD phase we adopted a supercell based on the 
a$_{\mathrm AFD}\sqrt{2}\times{{\mathrm a}_{\mathrm AFD}}\sqrt{2}\times{{\mathrm c}_{\mathrm AFD}}$.
The data are aligned with respect to the electron density (1 Nb per SC). 
Interpolating the polaron formation energy for $N \rightarrow \infty$
as a function of of $1/N^{3}$\cite{Makov} confirm the delocalization solution in the dilute limit.
The dotted line serves as a guide to the eye.
}
\end{figure}

\begin{figure}
\includegraphics[width=0.80\columnwidth,scale=1.0]{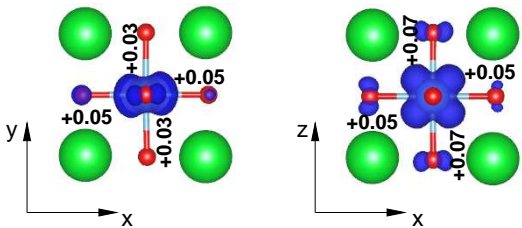}
\caption{\label{FIG:2} Local structural distortions of the Nb-induced polaron shonw in
Fig \ref{FIG:1}(a) within the $xy$ and $xz$ planes.
The numbers indicate the expansion (in \AA) with respect to the ideal cubic Ti-O bondlength.
}
\end{figure}

The substitution of one Ti with one Nb in a cubic $N=3$ SC,
corresponding to an electron density $n$ of $6 \times 10^{20}$ cm$^{-3}$, establishes a non-conducting state
with a rather sharp peak at about 0.80 eV below the conduction band minimum (CBM, see Fig. \ref{FIG:1}), 
in line with  the experimental values of 1.0 - 1.5 eV \cite{Fujimori, Fujimori2, Higuchi, Angelo, Yamada}. 
From an analysis of the charge character we infer that this state locates almost completely on one distinct 
Ti atom, 
and gives rise to a $d_{xz}^1$ configuration with a magnetic moment of about 0.9 $\mu_{\rm B}$. 
Consequentially, the Ti atom is reduced to Ti$^{3+}$. The electron trapping is accompanied by local
structural deformations manifested by the elongation of all Ti-O bond lengths within the Ti$^{3+}$O$_6$ octahedron
by about 0.05 \AA\, as shown in Fig. \ref{FIG:2}. This is the typical fingerprint of a small polaron.

The stability of the small polaron is determined by the competition between the structural energy 
required to distort the lattice $E_\mathrm{ST}$, and the electronic energy gained by localizing the excess 
electron at a specific Ti site $E_\mathrm{EL}$. 
We found a positive polaron formation energy of 57 meV for the smallest $N=3$ cubic supercell, 
which means that the small polaron is stable.

A key parameter, which has been largely overlooked in previous theoretical studies, is the role of
the carrier density. To clarify this issue, we have performed analogous calculations
using larger SCs, thus varying the doping concentration $x$ from $x$=3.7\% (N=3, $n$=6$\times$10$^{20}$ cm$^{-3}$) 
to $x$=0.5 \% (N=6, $n$=7$\times$10$^{19}$ cm$^{-3}$). 
We find that the polaron formation energy strongly depends on the carrier density, see Fig.~\ref{FIG:1}(c). 
For the smallest Nb-doping SC
(N=3) we have also studied the effects of the change of the volume due to Nb-doping. We found
that the lattice constant expands from a = 3.945 \AA\, to 3.993 \AA\, and the polaron formation energy increases up
to 140 meV. Following Vegard's law the volume expansion and its effect on the polaron formation energy
becomes negligible for larger supercell for which the Nb concentration is very small.

Within the low-$T$ AFD setup the small-polaron formation energy substantially increases for all values of
the carrier density [Fig.~\ref{FIG:1}(c), square symbols] using the optimized DFT+U volume.
This is explained by the larger Ti-O bond lengths in the low-temperature AFD structure as well as its higher 
structural flexibility compared to the cubic phase; both make the small polaron more favorable.
DFT-based calculations strongly overestimate the AFD distortions however \cite{Uchida, wahl}. 
Therefore, more realistic values for the AFD phase will be closer to those for the cubic phase. 
Indeed, by performing AFD calculations fro the N=4 SC at the experimental lattice constants the polaron formation energy 
is reduced by about 90 meV as indicated in Fig.~\ref{FIG:1}(c). 

The stability of the small polaron decreases as the carrier density decreases.  
For N$\approx$6 it eventually becomes negative, indicating that the excess electron
is prone to delocalization over the entire SC, or to create a large polaron. 

This localized-to-delocalized transition is explained by analyzing the energy balance involved in the electron trapping process.
The small polaron is stable when the electronic energy gain $|E_\mathrm{EL}|$ is larger than the structural 
energy loss $|E_\mathrm{ST}|$.
$E_\mathrm{ST}$ does not depend on the carrier density, but may depend on the presence of the dopant (Nb).
Indeed, we found that self-trapping in a perfect (defect free) lattice is unstable.
An increased Ti$^{3+}$-O bond length for the O atom binding to a neighboring Nb dopant indicate that the dopant provides
the crystal with the structural flexibility (i.e., small $\mathrm E_{ST}$)
needed to accommodate an electron in a locally distorted Ti$^{3+}$ site close to the Nb impurity.
This argument applies to other types of structural deformations such as surface reconstructions and steps,
which are found to favour small polaron formation. \cite{Setvin1, Setvin2}

We have verified that the stability of a polaron decreases by increasing the Ti$^{3+}$-Nb distance
and that the small polaron solution becomes unstable for Ti sites far from the Nb dopant.
This indicates that dopants are necessary to activate electron trapping:
without the dopant, excess electrons prefer to behave like free carriers. 
On the other hand, $E_\mathrm{EL}$ depends on the carrier density $n$ through electronic correlation.
The repulsion among the Nb-induced extra electrons is expected to get weaker at low 
concentration, attenuating the attraction between the excess electrons and the positively charge Nb center. 
Although small, this effect may become relevant for cases in which the polaron energy is very small
($E_\mathrm{ST}$ $\approx$ $E_\mathrm{EL}$).  

To ascertain the role of $U$ on the polaron formation energy we have also performed some 
calculations for different values of $U$ ($\pm$ 1 eV with respect to the cRPA value).
In agreement with previous studies we found that the polaron formation energy scales almost linearly 
with $U$\cite{Setvin1}: for $N$=4 the polaron formation energy becomes negative for $U-J=4$ (cubic) 
and $U-J=4.3$ (AFD).

Thus, in the limit of low concentration, $E_\mathrm{ST}$ becomes larger than $E_\mathrm{EL}$, the polaron
solution unstable and the excess electron begin to delocalize. 
It is extremely challenging to accurately describe delocalized large polarons from first principles. 
By incorporating many-body effects in  Fr$\ddot{\mathrm o}$hlich's large polaron Hamiltonian \cite{Frohlich},
Devreese {\em et al.}  \cite{Devreese} have computed the low-T optical conductivity of Nb-doped SrTiO$_3$ 
with large polarons for different dopant concentrations and found the best agreement with experiment at 
low carrier concentration ($n=1.7-3.4{\times}10^{19}$ cm$^{-3}$), 
in good agreement with the regime where we predict small polarons to be unstable [Fig.~\ref{FIG:1}(c)].
Inserting the characteristic electron-phonon coupling 
constant \cite{Devreese} and the characteristic longitudinal optical phonon frequency \cite{Servoin} in the 
Fr$\ddot{\mathrm o}$hlich's model, we obtain a spatial extension of the large polaron in SrTiO$_3$  of 9-19 \AA, {\em i.e.}, 3-5 unit cells.

Our findings resolve  puzzling and apparently conflicting results observed in Nb-doped SrTiO$_3$. 
First, the unusually narrow Drude feature for carrier concentrations between 0.1\% and 2\% per unit cell \cite{Mazin} 
can be understood by the large-polaron model, in line with our calculations for a small carrier density.
Second, the in-gap states detected by photoemission spectroscopy in moderately doped samples 
(1\% $<x<$ 5\%), suggest a semiconducting regime \cite{Higuchi}, in excellent agreement 
with our small polaron picture. 
At the transition from localized to free-carrier behavior, where the formation energy of small polarons is very small,
the thermal excitation is high enough to surmount it, resulting in a situation where the trapped and mobile 
electrons can coexist in the system. This picture can explain the existence of in-gap states in the
conductive phase measured by photoemission spectroscopy studies \cite{Fujimori, Fujimori2, Higuchi, Angelo},
and can be rationalized in terms of an inhomogeneous distribution of impurities or defects as suggested in
experimental works \cite{Szot, Muller}.

\subsection{Oxygen vacancies}
\begin{figure}
\includegraphics[width=0.90\columnwidth,scale=1.00]{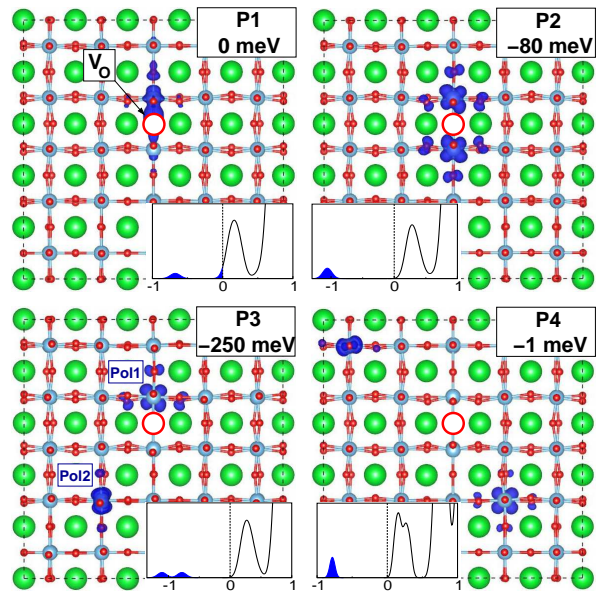}
\caption{\label{FIG:3}
(color online) Isosurfaces of the spin density (0.045 $e/\text{\AA}^2$) for one V$_\mathrm{O}$ (circle) in an N=5 SC. 
The insets show the density of states near the Fermi level (set at the CBM).
 The filled peaks indicate the polaron
states. In the P1 polaron configuration, considered to be the ground state in previous computational
studies \cite{Hou, Mitra, Choi, El-Mellouhi}, one electron 
is self-trapped near the V$_\mathrm{O}$, while the other one is delocalized. For each polaronic 
configuration (P1-P4) the relative energy with respect to the P1 phase is given (in meV). 
The most stable polaron solution is P3. The two small polarons (high spin density) are labeled Pol1 and Pol2.
A three dimensional view of the four polaron configurations is given in Fig. \ref{FIG:4}.
}
\end{figure}

To analyze the reduced sample we employed a cubic 5$\times$5$\times$5 SC with one V$_\mathrm{O}$,
resulting in two excess electrons per SC.
Previous computational analyses suggest that                    
one V$_\mathrm{O}$ induces a singly-occupied in-gap state, and donates one free carrier to the conduction 
band \cite{Hou, Mitra, Choi, El-Mellouhi, Janotti14}. We label this mixed localized/free-carrier polaron configuration P1.
The corresponding charge density isosurface is plotted in Fig.~\ref{FIG:3}.

\begin{figure*}
\includegraphics[scale=0.50]{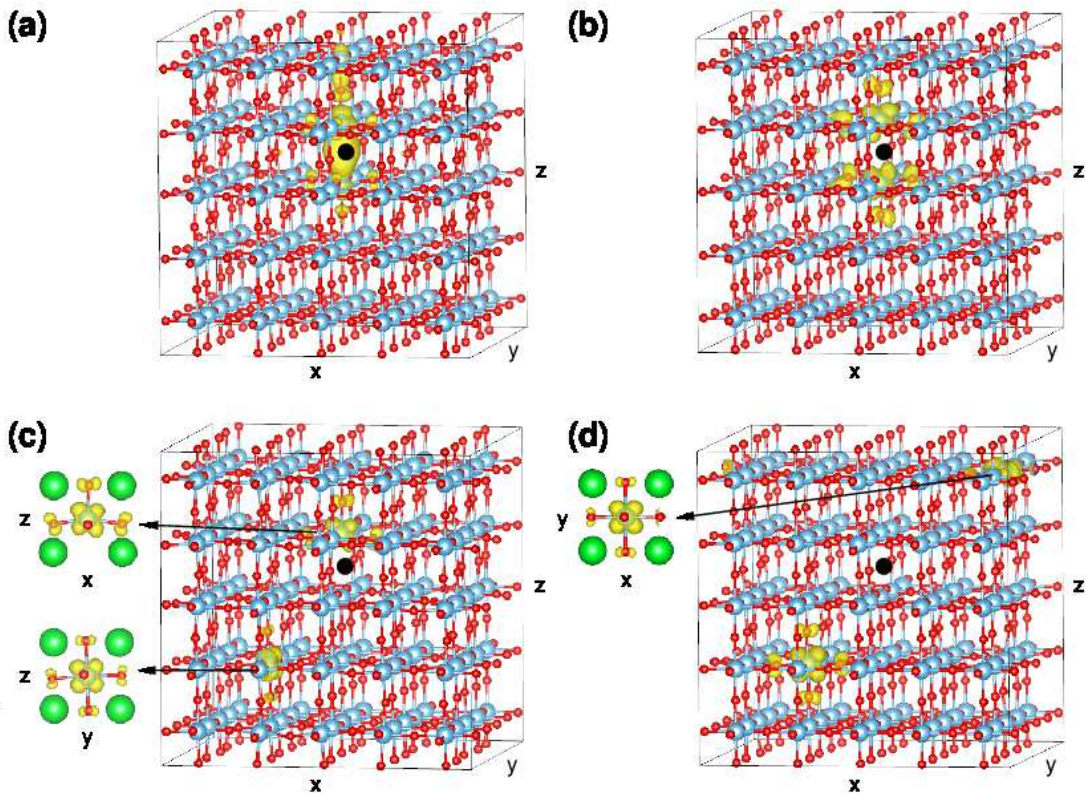}
\caption{\label{FIG:4}
(color online) Three dimensional view of the four polaron configurations shown in Fig. \ref{FIG:3}
and corresponding spin density isosurfaces. 
Small (red) and large (blue) spheres represent oxygen and titanium atoms, respectively.
Strontium atoms are not shown. The V$_\mathrm{O}$ is indicated by a (black) filled circle.
(a) P1: one electron is trapped near the V$_\mathrm{O}$ in a {\em d$_{z^2}$} orbital, while the second 
one is delocalized; 
(b) P2: both electrons are trapped near the V$_\mathrm{O}$ in {\em d$_{xz}$} orbitals;
(c) P3: one electron is trapped near the V$_\mathrm{O}$ in a {\em d$_{xz}$} orbital, while the second
one is located in a Ti$^{3+}$ site $\approx$ 8 \text{\AA} away with a {\em d$_{yz}$} configuration
(shown in the insets).
(d) both electrons are about 8 \AA\, away from the V$_\mathrm{O}$, and adopt {\em d$_{xz}$} and
{\em d$_{xy}$} configurations.
}
\end{figure*}

Based on a systematic search of several configurations we have identified
three more stable polaron patterns (P2, P3, and P4, Fig. \ref{FIG:3}).
All of them have an insulating density of states
with polaron peaks between 0.8 and 1.1 eV below the CBM. The polaron topologies of the various configurations 
differ in distance between the polaron and the V$_\mathrm{O}$, as depicted in Fig.~\ref{FIG:3} and \ref{FIG:4}.
The most stable configuration is P3, with one polaron anchored near V$_\mathrm{O}$ and the second one localized
in a Ti$^{3+}$ site $\approx$ 8 \AA~ away from V$_\mathrm{O}$. P3 is more favorable than P1 by 250 meV.
Different from previous calculations, all our additional polaron states display a {\em t$_{2g}$} distribution
 instead of {\em{ e$_{g}$}} ({\em d$_{z^2}$}).

From a structural point of view, removal of a single O atom makes the nearest-neighbor (NN) Ti (two) 
and Sr (four) atoms  move away from the 
vacant site, while the eight NN oxygen atoms relax toward the vacancy site, as already noted in Refs. \cite{Buban, Carrasco}.
In addition, the TiO$_6$ octahedra near V$_\mathrm{O}$ are subjected to AFD-like rotations \cite{Choi, Uchida}
(see Fig.~\ref{FIG:3}). 
These structural changes occur 
irrespective of the polaronic configuration. At the polarons (Ti$^{3+}$), the lattice relaxes similarly to the Nb-doped case.
We emphasize that a proper treatment of  local structural relaxations within a large unit cell is of 
primary importance for achieving a localized solution. 

Thus, a key element of our results is that the excess electrons induced by V$_\mathrm{O}$ 
do not form a mixed state of free and trapped carriers, as suggested by previous theoretical studies, 
but rather a "double polaron" with two trapped electrons in distinct Ti$^{3+}$ sites.

Now, how can this scenario explain the temperature-driven MIT observed in SrTiO$_{3-\delta}$ with homogeneously distributed 
V$_\mathrm{O}$s?
To answer this question we have conducted MD runs to probe the possibility to thermally activate the 
hopping mobility of the V$_\mathrm{O}$-induced small polarons.
We have monitored the position of the two polarons as a function of time
for different temperatures, each time starting from the ground state configuration P3.

\begin{figure}[h]
\includegraphics[clip,width=1.00\columnwidth,scale=1.00]{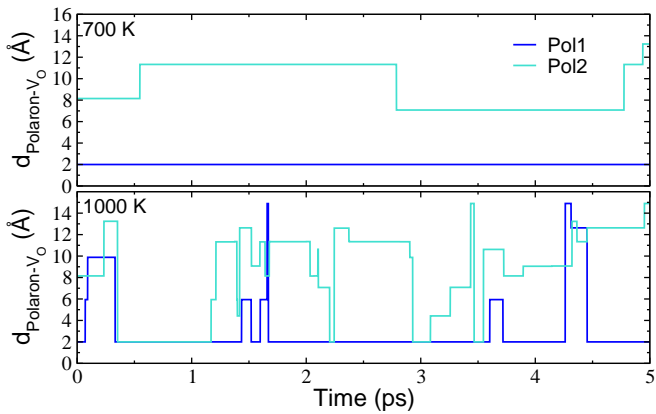}
\caption{\label{FIG:5}
(color online) MD in SrTiO$_{3-\delta}$, $\delta = 0.00267$ ($N=5$).
Mobility of the two polarons Pol1 and Pol2 starting from the P3 configuration (see
Fig. \ref{FIG:3}: Pol1 near V$_\mathrm{O}$, Pol2 about 8 \AA ~apart), given with respect to the distance
$d_{Pol1/Pol2-V_\mathrm{O}}$ between the polaron and V$_\mathrm{O}$ for two simulation temperatures, 700 K
(up) and 1000 K (bottom).   One polaron usually remains in the vicinity of V$_\mathrm{O}$, whereas the other one
frequently hops to neighboring sites.
}
\end{figure}

From the $T=700$\,K and $T=1000$\,K results in Fig. \ref{FIG:5} we infer that one of the polarons stays bound to 
the V$_\mathrm{O}$ forming a complex \cite{Janotti}.  This polaron gives rise to an optical absorption peak at 
about 1.1 eV below the CBM (see P3 DOS in Fig. \ref{FIG:3}), whereas the second one
can easily hop to other sites, enabling electrical conductivity.  
This polaron forms a peak at 0.8 eV below the CBM (see P3 DOS in Fig. \ref{FIG:3}). 
The oxygen vacancy does not undergo any diffusion during the MD run, in line with the reported
large migration barrier for oxygen diffusion (0.98 eV). \cite{Yamaji}

The mobility increases with increasing temperature.  At $1000$\,K we register 36 polaron migrations in 5 ps.
MD finds several other polaron topologies, with many minima close together on the potential energy landscape.
The P3 pattern remains the most favorable one ($\approx$ 65 \% of the time), followed by P2 
($\approx$ 20 \%) and P4 ($\approx$ 10 \%).
Importantly, MD never leads to a mixed localized/free-carrier P1-like configuration, although a degree of 
delocalization naturally emerges in the transition state from one minimum to the next one.   
The various minima could be populated at finite temperatures. Employing a simple Arrhenius equation, 
an activation energy of approximately 150 meV is estimated.\cite{note}

In order to assess the minimum energy barrier for small polaron hopping, we have 
investigated an abbreviated version of this process through 
the climbing image nudged-elastic band method \cite{Henkelman}, using  a
3 $\times$ 3 $\times$ 3 SC. We obtained migration barriers of 204 meV, 177 meV 
and 176 meV for small polaron hopping along the [100], [110] and [111] directions,
respectively, in agreement with the value derived from the MD runs.    

Our MD results provide a sound interpretation of the temperature-driven MIT in reduced SrTiO$_3$:
at high temperature the electrons exhibit hopping mobility, but once the temperature decreases, they will 
get trapped in polaron levels giving rise to a semiconducting regime\cite{Liu}. 

Finally, we would like to mention that we did not find any significant energy difference between the ferromagnetic 
and antiferromagentic arrangement of the trapped electrons. Hence, the hypothesis that V$_\mathrm{O}$s are 
 the cause of the observed ferromagnetism in SrTiO$_3$ \cite{Liu2} deserves further investigation. 

\section{Summary}

In summary, DFT+U calculations in conjunction with molecular dynamics simulations provide insights
into the behavior of excess electrons in SrTiO$_3$. We predict a carrier-density dependent transition between 
a semiconducting behavior with in-gap polaron states and a conductive regime.
Our findings reconcile 
conflicting experimental data and elucidates the fundamental physics behind. 
Our account of the oxygen defective SrTiO$_3$ at sufficiently high carrier concentration revisits previous theoretical results 
and leads to a unified and consistent explanation of the measured data: at low temperature the excess electrons are confined 
in Ti$^{3+}$ sites and become mobile when thermally activated.

\section{ACKNOWLEDGMENTS}
This work was supported by the ERC Advanced Research Grant `OxideSurfaces',
by the Austrian Science Fund (FWF) project F45, and by  the National Natural
Science Foundation of China (Grant No. 21201148 and 21303156). X.H. thanks
Zhicheng Zhong for helpful discussions.


\begin{thebibliography}{}
\bibitem{note}
In Ref. \cite{Liu}, an Arrhenius plot of the carrier density is interpreted as
an activation energy of 25\,meV, but this is for temperatures in the
`metallic' conductance regime. In the dopant model suggested there,
charge carrier saturation would occur under these conditions; thus the
interpretation of the Arrhenius plot is questionable.
\end{thebibliography}
\end{document}